\documentclass[11pt]{article}
\usepackage{moriond,epsfig}

\def\Journal#1#2#3#4{{#1} {\bf #2}, #3 (#4)}

\def\NPB{{\em Nucl. Phys.} B}
\def\PLB{{\em Phys. Lett.}  B}

\def\PRD{{\em Phys. Rev.} D}

\def\ra{\rightarrow}

\def\be{\begin{equation}}
\def\ee{\end{equation}}
\def\bea{\begin{eqnarray}}
\def\eea{\end{eqnarray}}

\def\ra{\rightarrow }

\def\epem{\mbox{e}^+\mbox{e}^- }

\def\kos{\mbox{K}^0_S }

\begin{document}
\vspace*{4cm}
\title{EXCLUSIVE CHANNELS IN PHOTON-PHOTON COLLISIONS AT LEP}

\author{S. BRACCINI}

\address{
INFN, Laboratori Nazionali di Frascati  \\ 
Via E. Fermi 40 - 00044 Frascati (Italy) \\
E-mail: Saverio.Braccini@lnf.infn.it    
}

\maketitle\abstracts{
The study of exclusive channels in photon-photon collisions
at e$^+$e$^-$ colliders allows to investigate the structure and the
properties of hadrons in a very clean experimental environment.
A concise review of the most recent results obtained at LEP is presented.
}

\section{Introduction}

 Photon-photon collisions at e$^+$e$^-$ colliders are studied using  the two-photon fusion
process $\epem\ra\epem{\rm X}$. In this reaction 
the outgoing electron and positron are usually scattered at very
small angles and are not detected. The two photons are quasi real
and the final state X must be neutral and unflavoured with C=1 and
J$\neq$1. The study of exclusive channels is characterized by a full reconstruction
of the final state X.

 The cross section for this process is given by the convolution of the QED calculable 
luminosity function $\cal{L}$, giving the flux of the virtual photons, with the two-photon cross section
$\sigma(\gamma\gamma\ra{\rm X})$ which is sensitive to the quark structure of the final state X. 

 The study of baryon-antibaryon pair production allows to probe the quark structure
of the baryons which can be modeled in terms of
three-quark~\cite{three-quark} and quark-diquark bound states~\cite{quark-diquark}.
Since the electric charge of the constituent partons is different in the two models,
the predictions for the cross section differ of more than one order of magnitude.

 If the final state is a single resonance R, $\sigma(\gamma\gamma\ra{\rm R})$ is expressed by
a Breit-Wigner function proportional to the two-photon width
$\Gamma_{\gamma \gamma}(\mbox{R})$ which contains all the physical information of the process. 
Below the charmonium threshold, the measurement of the two-photon width allows
to study QCD in the non-perturbative region and gives 
fundamental information on the nature of glueball candidates. 
Since gluons do not couple directly to photons, 
the two photon width of a glueball is expected to be
very small. Above the charmonium threshold,
the measurement of the two photon width of charmonia and bottomonia allows to
test perturbative QCD predictions. 

 If one of the two photons is highly virtual, the spin of
the final state is allowed to be one. The virtuality  
is taken into account in the cross section
by multiplying the Breit-Wigner function by a VDM pole
transition form factor which can be calculated using phenomenological models~\cite{Schuler,Cahn}.

 A more exhaustive review of the results obtained in this field in the last few years
can be found elsewhere~\cite{Saverio-Meson2000}.

\section{Baryon-antibaryon production}

\begin{figure}[t]
\begin{center}
\psfig{figure=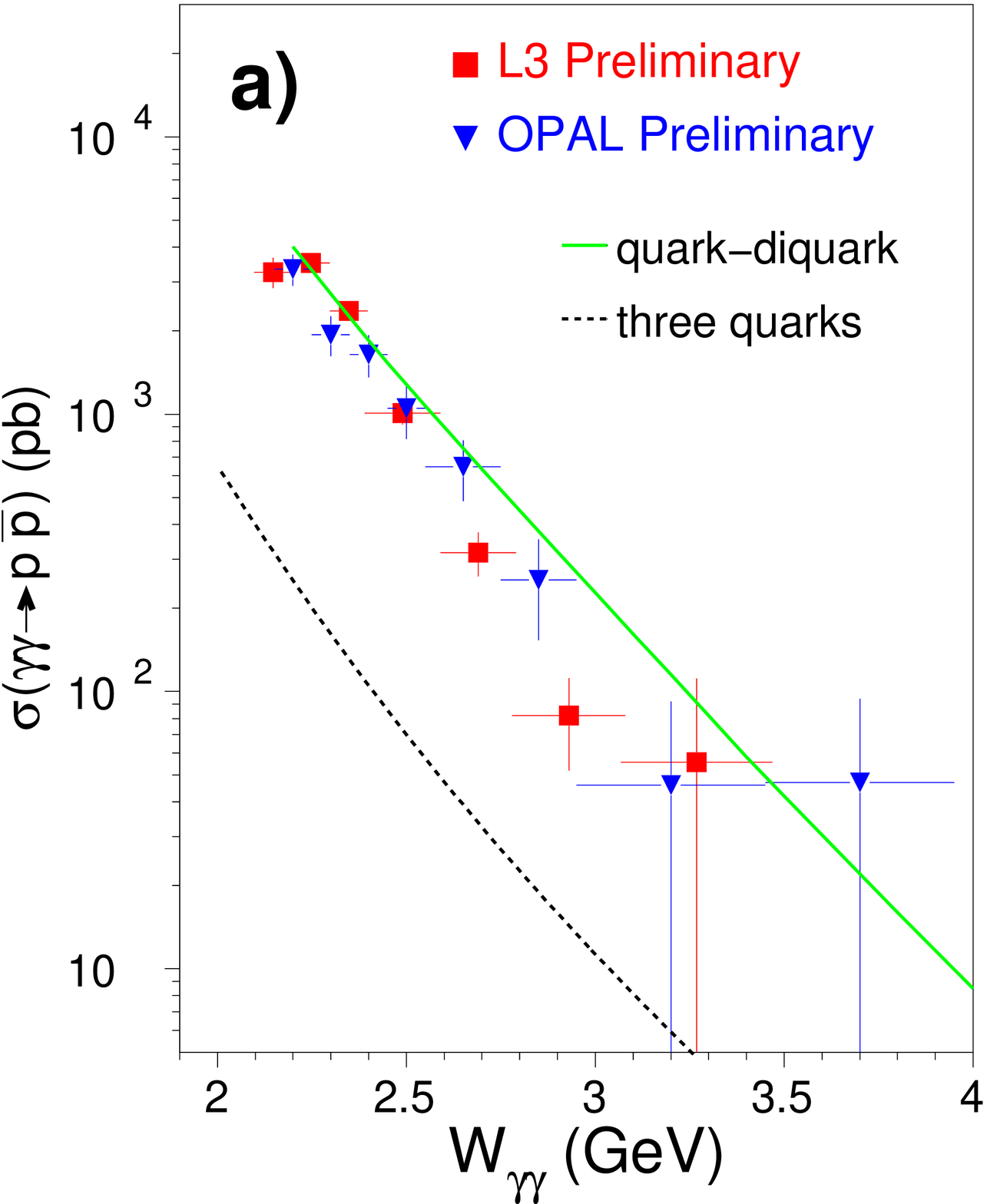,width=.3\textwidth}
\psfig{figure=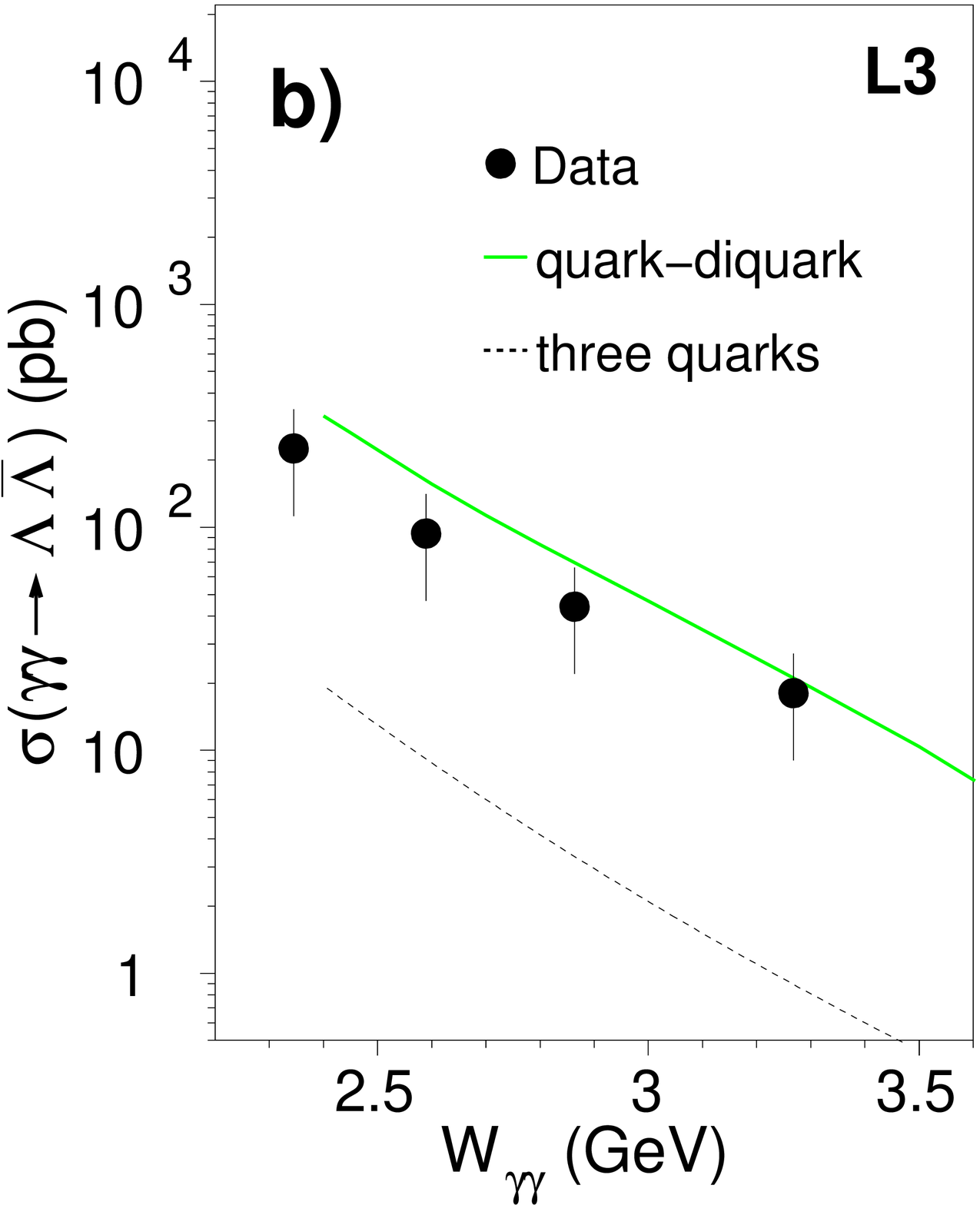,width=.3\textwidth}
\psfig{figure=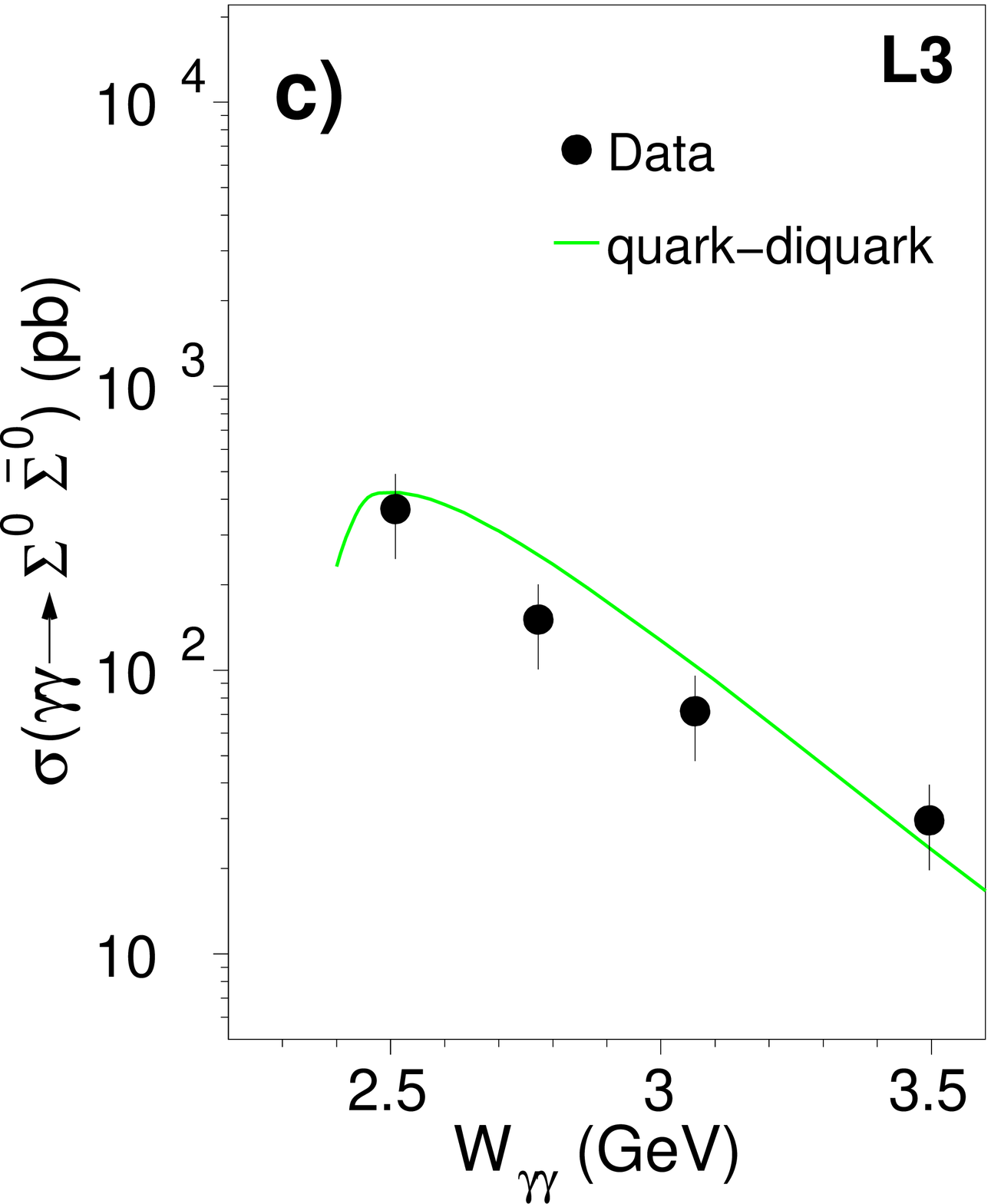,width=.3\textwidth}
\end{center}
\caption{The cross section for ${\rm p}\bar{{\rm p}}$ (a), 
$\Lambda\bar{\Lambda}$ (b) and $\Sigma^0\bar{\Sigma^0}$ (c) exclusive production.
\label{fig:barbabar}}
\end{figure}

 The cross section for ${\rm p}\bar{{\rm p}}$ exclusive production is measured by OPAL~\cite{OPAL-pp}
and L3~\cite{L3-pp} as a function of the two-photon effective mass $W_{\gamma\gamma}$,
using 249 pb$^{-1}$ and 610 pb$^{-1}$ 
of data collected at $\sqrt{s}=$ 183-189 GeV and at $\sqrt{s}=$ 183-209 GeV, respectively.
Good agreement is found between the two measurements and the predictions of the 
quark-diquark model~\cite{quark-diquark}, as presented in fig.~\ref{fig:barbabar}(a).
The predictions of the three-quark model~\cite{three-quark} are found to be inconsistent with the data.

 The study of the $\Lambda\bar{\Lambda}$ and $\Sigma^0\bar{\Sigma^0}$ exclusive production is performed
by L3~\cite{L3-llbar}. In particular, the $\Sigma^0\bar{\Sigma^0}$ exclusive production 
is studied for the first time.
Using 844 pb$^{-1}$ of data collected at $\sqrt{s}=$ 91-208 GeV, 19
$\Lambda\bar{\Lambda}$ and 14 $\Sigma^0\bar{\Sigma^0}$ candidate events are selected. The two measurements
of the cross section as a function of the two-photon effective mass $W_{\gamma\gamma}$
are shown in fig.~\ref{fig:barbabar}(b) and fig.~\ref{fig:barbabar}(c).
Good agreement is again found with the predictions of the quark-diquark model. Data are inconsistent
with the three-quark model predictions.

\section{Formation of light resonances}

\begin{figure}[t]
\begin{center}
\psfig{figure=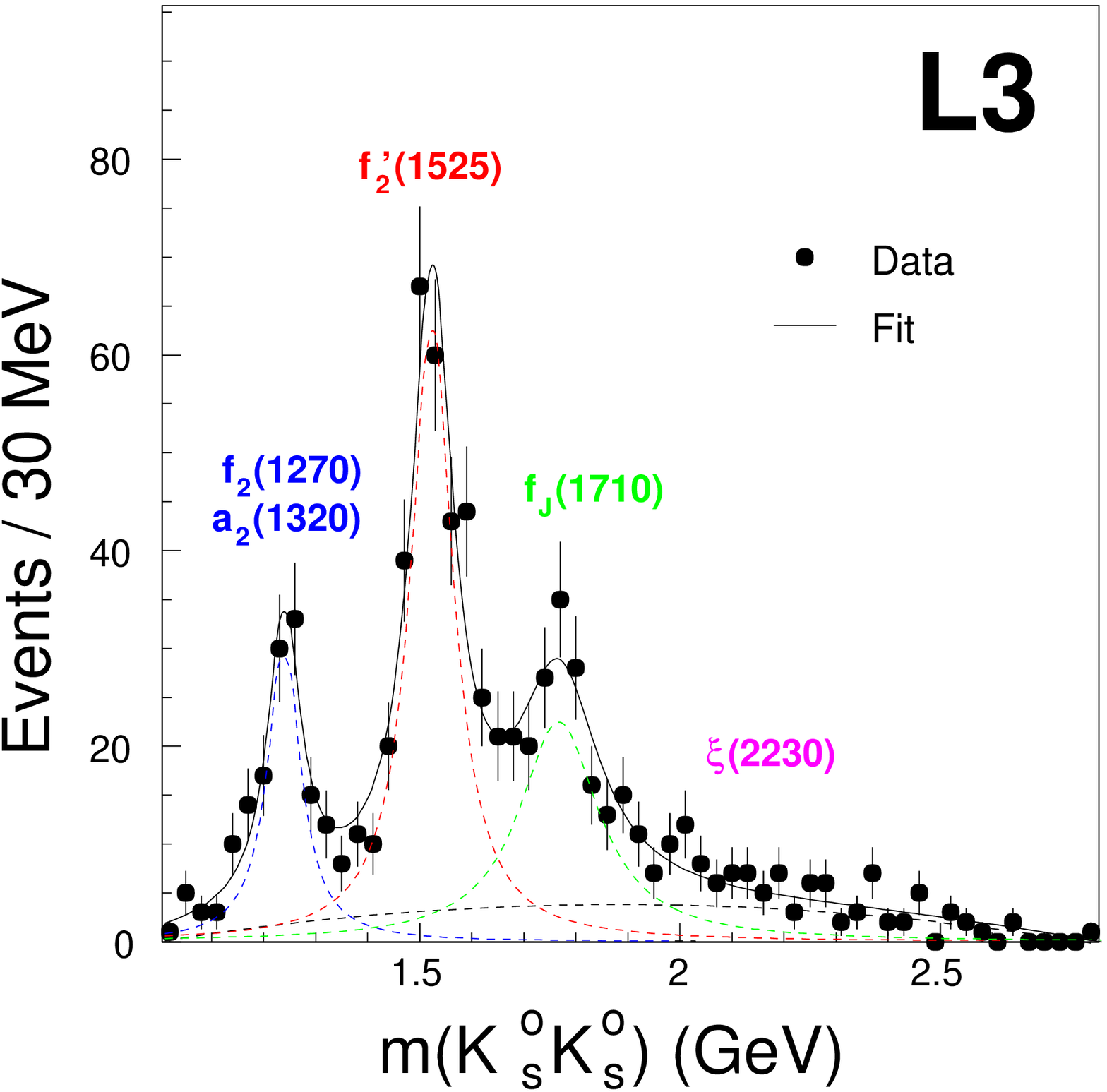,width=0.42\textwidth}
\hspace*{4mm}
\psfig{figure=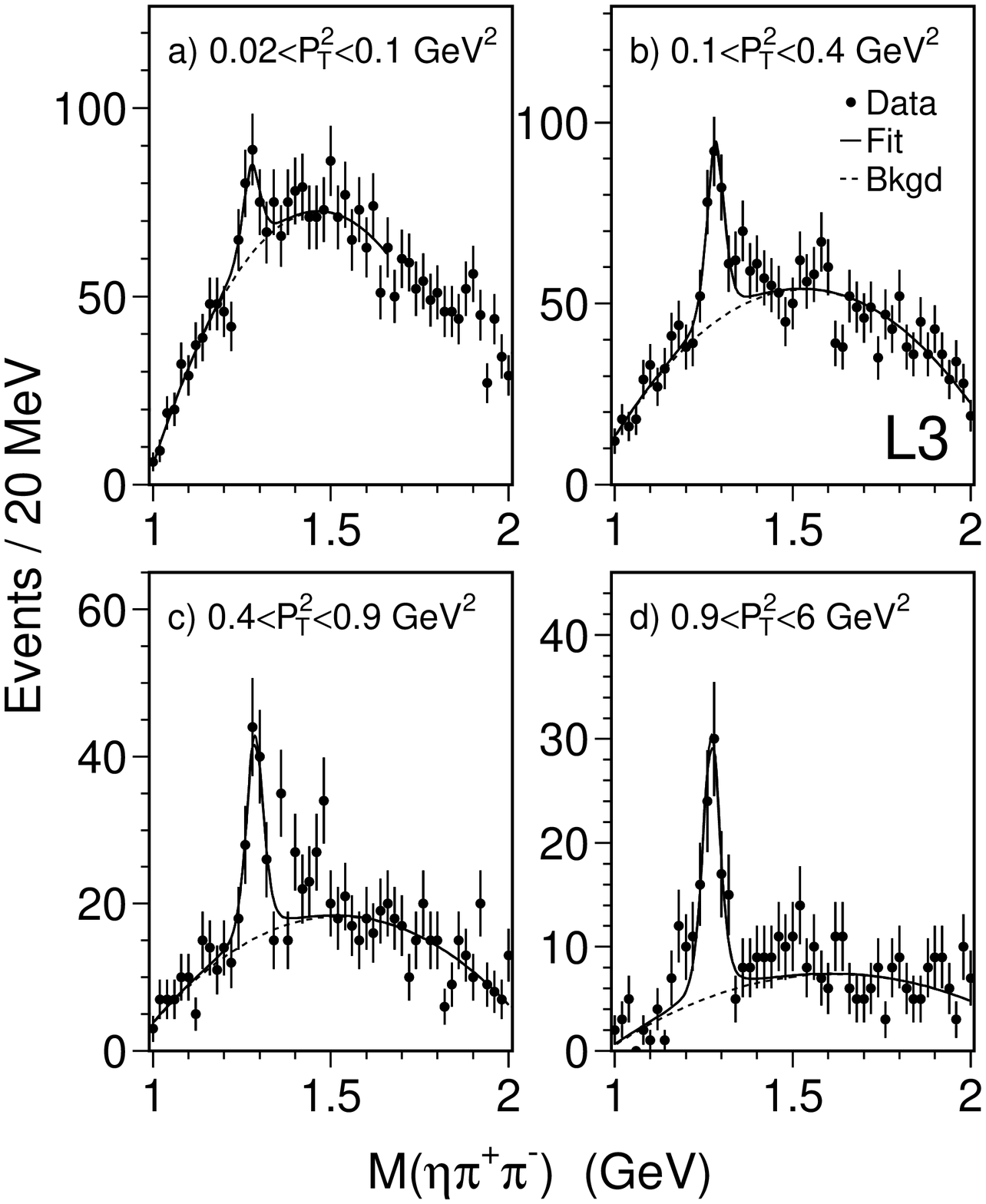,width=0.36\textwidth}
\end{center}
\caption{The $\kos\kos$ mass spectrum (left) and the $\eta\pi^+\pi^-$ mass
spectrum as a function of $Q^2$ (right).
\label{fig:resonance}}
\end{figure}

 The $\kos\kos$ final state is studied by L3~\cite{L3-kkbar}
using 588 pb$^{-1}$ of data collected at $\sqrt{s}=$ 91-202 GeV.
The $\kos\kos$ mass spectrum is found to be
characterized by three resonant signals over a small background,
as presented in fig.~\ref{fig:resonance}(left). 
The peak in the
1100$-$1400 MeV mass region is due to the
destructive f$_2$(1270)$-$a$_2$(1320) interference. 
The spectrum is dominated by the formation of the f$_2\,\!\!\!'$(1525) tensor meson
in helicity 2 state for which
$\Gamma_{\gamma\gamma}({\rm f}_2'(1525))\times \mbox{BR}({\rm f}_2'(1525)\rightarrow
\mbox{K}\bar{\mbox{K}})$= 76 $\pm$ 6 $\pm$ 11 eV. 
The signal at 1750 MeV mass is found to be dominated by the 2$^{++}$, helicity 2 wave.
This is interpreted as
the formation of a radially excited tensor meson for which
$\Gamma_{\gamma\gamma}({\rm f}_2(1750))\times \mbox{BR}({\rm f}_2(1750)\rightarrow
\mbox{K}\bar{\mbox{K}})$= 49 $\pm$ 11 $\pm$ 13 eV.
A fraction of 24 $\pm$ 16\% of the 0$^{++}$ wave is also found.
No signal for the 
$\xi$(2230) tensor glueball candidate is observed. 
The upper limit $\Gamma_{\gamma\gamma}(\xi(2230))\times $BR$(\xi(2230)\ra\kos\kos)<1.4$ eV 
at 95\% C.L. is obtained.

 The formation of spin one states in two-photon collisions is possible if one of the two photons
is virtual. The virtuality $Q^2$ is evaluated experimentally by measuring the total transverse
momentum imbalance $P_t$ of the decay products of the resonant state. Monte Carlo studies
show that  $Q^2\simeq P_t^2$. The $\eta\pi^+\pi^-$ final state as a function
of $Q^2$ is studied by L3~\cite{L3-f1}. The four mass spectra presented in 
fig.~\ref{fig:resonance}(right) show the evolution of the formation of the f$_1$(1285)
as a function of $Q^2$. 
The signal reduces to zero at low $Q^2$, due to the Landau-Yang theorem which forbids 
the formation of spin-one states by two real photons.
From these mass spectra
the cross section is measured as a function of $Q^2$ and is compared to the theoretical
predictions based on different parametrizations of the transition from factor. The calculation 
by Cahn~\cite{Cahn} leads to a confidence level below $10^{-9}$ and is therefore found incompatible with
the data. The parametrization by Schuler et al.~\cite{Schuler} 
leads to a confidence level of 2\%
if the mass of the pole is fixed to the mass of the f$_1$(1285). A confidence level of
91\% is found leaving the parameters $\Lambda$ and $\tilde{\Gamma}_{\gamma\gamma}$ free in the fit.
From this fit, the values $\Lambda = 1.04 \pm 0.06 \pm 0.05$ GeV and 
$\tilde{\Gamma}_{\gamma\gamma} = 3.5 \pm 0.6 \pm 0.5$ keV are obtained.

\section{Search for the $\eta_b$}

 The $\eta_b$ meson is the ${\rm b}\bar{{\rm b}}$ ground state and its
experimental observation is still missing.
According to theoretical perturbative QCD and lattice QCD predictions,
the difference $m(\Upsilon)-m(\eta_b)$ is in the range from 34 to 141 MeV and the
two-photon width of the $\eta_b$ is between 500 and 570 MeV. The branching ratios
into four and six charged particles are estimated to be 2.7\% and 3.3\% respectively.
The search for the $\eta_b$ represents therefore an exciting challenge and a very
important test for QCD. 

 Due to the high mass of this state, the background from other two-photon processes 
is very small and high energy LEP data above the W pair production
threshold represent a very good environment to search for this meson. According
to the predictions, about six $\eta_b$ mesons are expected to be produced per
decay channel and per experiment. The number of observed events will be sensitively reduced 
due to acceptance and efficiency effects.

\begin{figure}[t]
\begin{center}
\psfig{figure=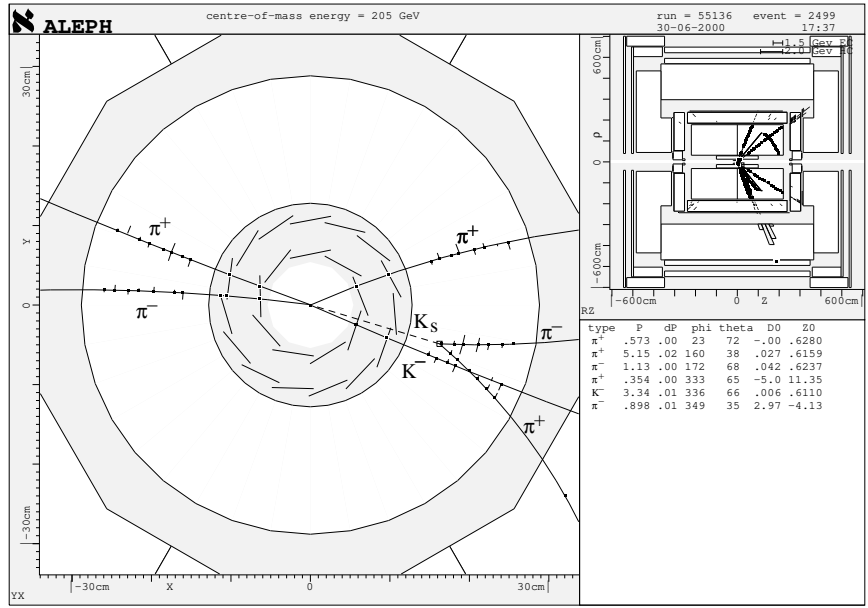,width=0.5\textwidth,angle=-90}
\end{center}
\caption{The $\eta_b\rightarrow$ 6 charged particles candidate event by ALEPH. 
\label{fig:etab}}
\end{figure}

 A search for the $\eta_b$ meson is performed by ALEPH~\cite{ALEPH-etab} using an integrated luminosity of
699 pb$^{-1}$ collected at  $\sqrt{s}=$ 181-209 GeV. No candidates are found in the
four charged particle decay mode. With an expected background of 0.30$\pm$0.25 events,
the upper limit  
$\Gamma_{\gamma\gamma}(\eta_b)\times\mbox{BR}({\rm 4\: charged}) <$ 48 eV
at 95\% C.L. is derived. One candidate is found in the six charged particle dacay mode.
This candidate is shown in fig.~\ref{fig:etab} and is found to have a mass of 
9.30 $\pm$ 0.02 $\pm$ 0.02 GeV, very close to theoretical predictions.
It is important to remark the clear presence of a $\kos$
and a  K$^-$ strange mesons in the event. Since the direct formation of s quarks in two-photon collisions
is suppressed, this event is very probably due to the formation of a high mass resonant state.
With an expected background of 0.70$\pm$0.34 events,
the upper limit  
$\Gamma_{\gamma\gamma}(\eta_b)\times\mbox{BR}({\rm 6\: charged}) <$ 132 eV
at 95\% C.L. is derived.

 In a preliminary study of the formation of the $\eta_b$ meson, L3~\cite{L3-etab} reports the
observation of some candidates in the mass region of interest. Using a luminosity of
610 pb$^{-1}$ collected at  $\sqrt{s}=$ 183-209 GeV, 1 candidate in the four
charged, 1 candidate in the six charged and
2 candidates in the two charged particle and one $\pi^0$ final states are found. 

 These results indicate a possible formation of the  $\eta_b$ meson in photon-photon fusion.
 Due to the very limited statistics, a combination of the results of all the four LEP collaborations
will be mandatory to have the possibility to claim the observation of this state.

\section*{Acknowledgments}

 I would like to acknowledge the LEP collaborations ALEPH, OPAL and L3 for having
provided me with their results. I would like to thank M.N. Kienzle-Focacci, B. Echenard,
I. Vodopianov, T. Barillari and A. Boherer for all the constructive discussions and suggestions.

\end{document}